\documentclass[12pt]{article}

\usepackage{amsfonts}
\usepackage{amssymb}
\usepackage{amsthm}
\usepackage[fleqn]{amsmath}
\usepackage[mathcal]{euscript}
\usepackage{mathrsfs}
\usepackage{graphicx}
\usepackage{float}

\graphicspath{{./Figures/}}

\renewcommand{\iint}{{\int\!\!\!\!\int}}
\renewcommand{\iiiint}{{\int\!\!\!\!\int\!\!\!\!\int\!\!\!\!\int}}

\newcommand{\Ref}[1]{(\ref{#1})}

\newcommand{\kB}{k_{\mathrm{B}}} 
\newcommand{\SB}{S_{\mathrm{B}}} 

\newcommand{\eps}{\epsilon}
\newcommand{\veps}{\varepsilon}

\newcommand\sett{{\mathfrak{s}}}

\newcommand{\aV}{\mathbf a}           
\newcommand{\pV}{\mathbf p}           
\newcommand{\qV}{\mathbf q}           
\newcommand{\sV}{\mathbf s}           
\newcommand{\uV}{\mathbf u}           
\newcommand{\vV}{\mathbf v}           

\newcommand{\LV}{\mathbf L}           
\newcommand{\lambdaV}{\boldsymbol{\lambda}}           

\DeclareMathAlphabet{\mathpzc}{OT1}{pzc}{m}{it}

\newcommand\pzcE{{\mathpzc{E}}}

\newcommand\pzcS{{\mathpzc{S}}}

\newcommand\pzcSB{\pzcS_{\mathrm{B}}}

\newcommand{\dd}{\mathrm{d}}

\newcommand{\Rset}{\mathbb{R}}
\newcommand{\Sset}{\mathbb{S}}

\newcommand{\Asp}{\mathfrak{A}}

\newcommand{\tst}{\textstyle}

\newcommand{\tfrhalf}{{\tst{\frac{1}{2}}}}

\newcommand{\tfrquart}{{\tst{\frac{1}{4}}}}

\newcommand{\chord}[1]{| #1 |}

\begin{document}

\title{The influence of gravity on the\\ Boltzmann entropy of a closed universe}

\vspace{-0.3cm}
\author{\normalsize \sc{Michael K.-H. Kiessling}\\[-0.1cm]
	\normalsize Department of Mathematics, Rutgers University\\[-0.1cm]
	\normalsize Piscataway NJ 08854, USA}
\vspace{-0.3cm}
\date{$\phantom{nix}$}
\maketitle
\vspace{-1.6cm}

\begin{abstract}
\noindent
  This contribution inquires into Clausius' proposal that ``the entropy of the world tends to a maximum.''
 The question is raised whether the entropy of `the world' actually does have a maximum; and if the answer is ``Yes!,'' 
what such states of maximum entropy look like, and if the answer is ``No!,'' what this could entail for the fate of the 
universe.
 Following R. Penrose, `the world' is modelled by a closed Friedman--Lema\^{\i}tre type universe, 
in which a three-dimensional spherical `space' is filled with `matter' consisting of $N$ point particles, 
their large-scale distribution being influenced by their own gravity. 
 As `entropy of matter' the Boltzmann entropy for a (semi-)classical macrostate, and Boltzmann's
ergodic ensemble formulation of it for an isolated thermal equilibrium state, are studied.
 Since the notion of a Boltzmann entropy is not restricted to classical non-relativistic physics,
the inquiry will take into account quantum theory as well as relativity theory; we also consider black hole entropy.
 Model universes having a maximum entropy state and those which don't will be encountered. 
 It is found that the answer to our maximum entropy question is not at all straightforward at the general-relativistic
level.
 In particular, it is shown that the increase in Bekenstein--Hawking entropy of general-relativistic black holes does 
not always compensate for the Boltzmann entropy of a piece of matter swallowed by a black hole.
\end{abstract}

\vfill
\hrule
\smallskip\noindent
{\small 
Typeset in \LaTeX\ by the author. Revised version of May 10, 2019. 

\smallskip\noindent
Invited paper for a volume on the philosophy of statistical mechanics, titled 
``Statistical Mechanics and Scientific Explanation: Determinism, Indeterminism, and the Laws of Nature,'' 
edited by  {\sl Valia Allori}.

\smallskip\noindent
\copyright 2019 The author.
This preprint may be reproduced for non-commercial purposes.}
\newpage
\section{Introduction}
\vskip-.2truecm
\noindent
        The notion of \textit{entropy}, roughly meaning ``amount of transformation,'' was introduced into science 
in 1865 by Rudolf Clausius in his path-breaking paper \cite{Clausius} on the mechanical theory of heat, where
he recast what was known as \textit{the first and second laws of the theory of heat } 
in a format still featuring prominently in today's thermodynamics.
  Although Clausius' discovery of the entropy concept was based on his sound mathematical reasoning, and on
his careful analysis of empirical evidence obtained in laboratory experiments, he was thinking in much bigger terms.
  Indeed, at the very end of his paper he argued that the fate of the whole universe is ruled by the two laws as follows:
\medskip

\hspace{1truecm}{Law 1: \textit{The energy of the world is constant.}}

\hspace{1truecm}{Law 2: \textit{The entropy of the world tends to a maximum.}}
\medskip

\noindent
In Law 2) it is tacitly understood that this tendency is monotonic.

 Since the fate of the whole universe is at stake it may be worthwhile to examine Clausius' bold proposal more closely. 
 In this contribution we ask: 
\medskip

{Q1: \textit{Does Clausius' world (or its kin) have an entropy maximum?}}

{Q2: \textit{If the answer to Q1 is ``Yes!'', what is its maximum entropy state?}}

{Q3: \textit{If the answer to Q1 is ``No!'', then what is the fate of the universe?}}
\medskip

 Since Clausius' notion of the world (universe) has long been superseded thanks to the revolutionary 
developments in observational, experimental, and theoretical physics, to avoid a pointless academic excercise
we will examine Clausius' proposal not merely from the perspective of classical Newtonian-type theory which ruled
supreme at Clausius' times, but we also take quantum theory and relativity theory into account. 
 Thus we use a sequence of increasingly more realistic model worlds with finite matter and energy content.
 No pretense is made that our findings correctly anticipate the fate of our real universe; yet, we hope that they offer
some insights into its inner workings.

 We follow R. Penrose~\cite{penroseR} and carry out our inquiry in the setting of a closed universe, defined in section 2
and Appendix A.
 In section 3 we recall Boltzmann's entropy.
 In section 4 we study the three-dimensional classical model with Newtonian gravity and find
the Boltzmann entropy to be without upper bound. 
  In an appendix to section 4 (Appendix B) we demonstrate that the
Boltzmann entropy would actually have a maximum if the Newtonian gravitional pair energy would only diverge
logarithmically with the separation of the particles rather than reciprocally;
if space were two-dimensional then a spherical Newtonian universe would be just like that.
 Section 5 deals with a non-relativistic quantum-mechanical improvement over the model of section 4.
 It is found that quantum mechanics stabilizes the divergent Newtonian gravitational attraction at short
distance, as a consequence of which the quantum analog of Boltzmann's entropy features a maximum. 
 In section 6 we take special relativity into account, while gravity is still Newtonian.
 Remarkably, special relativity destabilizes, and in a world containing more than about one solar mass the
quantum Boltzmann entropy does not have a maximum. 
 At long last, in section 7, we inquire into general-relativistic issues, and the role of the Bekenstein--Hawking
black hole entropy.
 Roger Penrose has argued that the Bekenstein--Hawking entropy of a black hole which has swallowed all the matter
is the maximum entropy state of a closed universe. 
 We dispute this proposal by showing that the sum of the Bekenstein--Hawking black hole entropy and the 
quantum Boltzmann entropy of the cosmic microwave background radiation (``matter'') outside the black hole's event horizon 
does not always increase if the black hole swallows up some of the radiation. 
 To rescue the second law we suggest that the (quantum) Boltzmann entropy of matter inside the horizon has to be taken into
account.

\section{The closed universe framework}\vspace{-5pt}

 We will for the most part ignore that our universe seems to have had a beginning in the ``Big Bang,''
and that it is expanding, and that it may well not be closed but open. 
 Here, by ``closed'' we mean ``of finite spatial extent,'' while ``open'' means ``unbounded,''
which is the usual terminology in cosmology. 
 Since an unbounded universe with an infinite, or finite, amount of matter in it would almost inevitably have no
upper limit to its entropy, and since this was almost surely clear to Clausius, it stands to reason that Clausius
had in mind a closed universe of finite matter content.
 We contemplate, following James Hopwood Jeans \cite{Jeans} (and subsequently Albert Einstein \cite{Einstein} and 
Roger Penrose~\cite{penroseR}), that space is spherical, i.e. a three-dimensional sphere of radius $R$, viz. ${\mathbb S}_R^3$.
 The easiest way to think about ${\mathbb S}_R^3$ is as the subset of Euclidean space vectors $\sV\in{\mathbb R}^4$ restricted
to have Euclidean length $|\sV| = R$.
 Writing $\sV\in{\mathbb S}_R^3$ means just this.
 We also assume that cosmological time $t\in\Rset$.
 
 We will for simplicity assume that matter in this universe consists of a single species of `fundamental' particles, 
which we call `atoms,' represented by $N$ point particles with gravitational interactions between them. 
 For most of the discussion we will use Newtonian gravity (see Appendix A),
with general relativity taken into account at a later stage.
  
 To characterize a microstate of $N$ Newtonian point particles it is not enough to give all their positions.
 One also needs their velocities, respectively their momenta.
 The notion of Newton's mechanics extended to particle motions in non-Euclidean spaces was worked out by
Killing, see \cite{Killing}, but in our simple geometrical setup we do not need to invoke the abstract formalism 
of differential geometry.
 True, since physical space in our spherical universe model is not flat but a Riemann manifold with constant curvature, 
we cannot simply add or subtract two points in $\Sset^3_R$ to get a new point in $\Sset^3_R$. 
 However, since we can think of points $\sV\in\Sset^3_R$ as vectors $\sV\in\Rset^4$ of length $R$, 
we can add or subtract such four-dimensional vectors to get a new four-dimensional vector in $\Rset^4$.
 In particular, we can define the particle velocities $\vV(t):=\frac{d}{dt}\qV(t)$ at $\qV(t)\in \Rset^4$ in the usual
vector calculus way as a limit when $\Delta t\to 0$ of the vector differential 
quotient $\frac{\qV(t+\Delta t) - \qV(t)}{\Delta t}$ in $\Rset^4$, and for 
motions which take place in the subset $\Sset^3_R\subset\Rset^4$ find that 
$\vV(t)$ is always tangential to $\Sset^3_R$ at $\qV(t)\in\Sset^3_R$.
 Thus, given $\qV(t)\in\Sset^3_R$ at time $t$, one only needs three numbers to characterize its 
velocity $\vV(t)$, a three-dimensional vector in the tangent space $T_{\qV}{\mathbb S}_R^3$ of the 
instantaneous particle position $\qV\in{\mathbb S}_R^3$. 
 Of course, for each $\qV$, the attached tangent space $T_{\qV}{\mathbb S}_R^3\sim{\mathbb R}^3$, 
but to compare vectors in tangent spaces for different $\qV$ and $\qV'$ one needs a \emph{connection}. 
 We don't need to worry
about this notion in its generality here because we can think of each  $T_{\qV}{\mathbb S}_R^3$ naturally
as a three-dimensional Euclidean subspace of ${\mathbb R}^4$ in which ${\mathbb S}_R^3$ is embedded.

 As for the particle momenta, in Newtonian physics they are simply the product of the particle velocities with their
inert mass $m$, yet because of the subsequent developments (Lagrange and Hamilton formulations) momenta are
nowadays considered to not live in the tangent space but in the co-tangent space $T^*_{\qV}(\Sset^3)\sim {\mathbb R}^3$ 
of the particle position $\qV\in {{\mathbb S}_R^3}$. 
 Elements of the co-tangent space act on elements of the tangent space as bounded linear 
maps into the real numbers. 
 In the Newtonian setting there is of course the obvious identification $\pV(\uV) = m\vV\cdot\uV$ 
(Euclidean inner product in ${\mathbb R}^3$), with both $\vV$ and $\uV$ velocities in the same $\Rset^3$.
 This is a fine point which is made for the sake of mathematical accuracy; it is not important
for the conceptual developments in this paper.

\section{Boltzmann's classical entropy formulas}\vspace{-5pt}

 We need a formula which allows us to study the entropy of such a closed universe.
 While Clausius was an atomist, 
he did not propose a formula for computing entropy based on a mechanical atomistic theory of matter. 
 This step had to wait until 1872 when Ludwig Boltzmann introduced his ``$H$ function'' (essentially the negative of
the entropy) into the kinetic theory of gases; cf. \cite{BoltzmannA}.
 We here adapt Boltzmann's $H$ function formula to our setting of a dilute, monatomic, ideal classical gas in 
${{\mathbb S}_R^3}$. 
 We also take into account the subsequent adjustments physicists have made to Boltzmann's entropy formula: 
we add Gibbs' \cite{Gibbs} combinatorial term $-\ln N!$ which accounts for the permutations of the $N$ 
identical particles, and we quantify the entropy in units of 
the Boltzmann constant $\kB$ and the phase space measure of a particle in units of 
the cube of Planck's constant\footnote{The Boltzmann constant $\kB$ and the Planck contant $h$ were 
  introduced by Max Planck.}
$h$.
 The Boltzmann entropy of such a classical gas reads
\begin{equation}\label{SB}
\pzcS_{\mathrm{B}}(f;N)
 = -\kB N\ln N -  \kB N \int_{{\mathbb S}_R^3}\Big(\int_{T^*_{\sV}(\Sset^3_R)} f \ln (h^3 f/e) \dd^3{p}\Big)\dd^3{s}\, ;
\end{equation}
here, $f(\sV,\pV,t)$ is a continuum approximation to the \emph{normalized empirical} (i.e. actual) density of the
$N$-atom gas in the six-dimensional co-tangent bundle\footnote{If instead of spherical space $\Sset^3_R$ we would
  work with flat space $\Rset^3$, the co-tangent spaces at different points $\qV_1$ and $\qV_2$ would just be 
  Euclidean translates of each other, and the co-tangent bundle of $\Rset^3$ would become just the Cartesian product
  $\Rset^3_{(q)}\times\Rset^3_{(p)}$, where the suffix ${}_{(q)}$, resp. ${}_{(p)}$ indicates position space, resp.
 momentum space.
  But $\Sset^3_R$ is not a linear space, so its co-tangent bundle $\bigcup_{\sV\in{\mathbb S}_R^3}T^*_{\sV}(\Sset^3_R)$ 
is a more complicated manifold. 
  Yet thanks to the embedding $\Sset^3_R\subset\Rset^4$, the bundle $\bigcup_{\sV\in{\mathbb S}_R^3}T^*_{\sV}(\Sset^3_R)$ 
is a six-dimensional subset of $\Rset^4_{(q)}\times\Rset^4_{(p)}$.} 
 $T^*(\Sset^3_R):= \bigcup_{\sV\in{\mathbb S}_R^3}T^*_{\sV}(\Sset^3_R)$ 
 of the physical space ${{\mathbb S}_R^3}$, at ``cosmic time'' $t$.

 We pause for a moment to comment on (\ref{SB}).

 In the physics literature, the co-tangent bundle  $T^*(\Sset^3_R)$
of physical space ${{\mathbb S}_R^3}$ is frequently called the ``one-particle phase space,'' but this 
terminology is misplaced in the context of (\ref{SB}). 
 Namely, in the interpretation of $T^*(\Sset^3_R)$ as one-particle phase space we would have
$T^*(\Sset^3_R) = \bigcup_{\qV\in{\mathbb S}_R^3}T^*_{\qV}(\Sset^3_R)$ (N.B.: $\qV$ denotes the position of a 
particle, whereas $\sV$ denotes a point in space irrespectively of whether that point $\sV$ is occupied by a point
particle or not), and then the non-negative function $f(\qV,\pV,t)$ would
not be a continuum approximation to an actual normalized density, but would instead have the meaning of a probability density 
for finding that single particle having position $\qV\in{{\mathbb S}_R^3}$ and momentum $\pV\in T^*_{\qV}(\Sset^3_R)$.
 Except for the factor $N$ and the additive $-N\ln N$, this would turn the integral in (\ref{SB}) into an expected 
value of $\ln f$ w.r.t. $f$, essentially. 
 In other words, the integral in (\ref{SB}) would be an ensemble entropy --- a Gibb's entropy for a single-particle ensemble with
ensemble probability density $f$.
 This would make its factor $N$ quite incomprehensible, not to speak of the $N\ln N$ term. 

 Clearly, thinking of the integral in (\ref{SB}) as an ensemble entropy
means to completely miss Boltzmann's point that (\ref{SB}) is the physical entropy of an 
individual $N$ body system, with $Nf(\sV,\pV,t)$ a continuum approximation to the particle density in the co-tangent bundle 
$\bigcup_{\sV\in{\mathbb S}_R^3}T^*_{\sV}(\Sset^3_R))$ of \emph{physical space} (here: ${{\mathbb S}_R^3}$).
  To obtain $f(\sV,\pV,t)$ at time $t$ from a system with $N\gg 1$ particles, for each $\sV\in{{\mathbb S}_R^3}$
consider a small sphere centered at $\sV$ containing itself $n\gg 1$ particles, but with $n\ll N$. 
 The momenta of the particles in this little sphere can be distributed into bins, forming a histogram over $T^*_{\sV}(\Sset^3_R)$
that is given a continuum approximation --- in the binning process the $\pV$ vectors at different points $\qV$
need to be compared with $\pV$ vectors at $\sV$ for which the earlier-mentioned connection is needed. 
 By contrast, no such connection would be needed if one would merely consider a theoretical probability of a
particle at a given $\qV$ having a momentum $\pV$. 

 Boltzmann's formula (\ref{SB}) results from Boltzmann's more general definition in \cite{BoltzmannB} of the entropy 
of a macrostate of an individual system, famously summarized by Max Planck as 
\begin{equation}\label{SlogW}
 S = \kB \log W,
\end{equation}
where $W$ (for the German ``Wahrscheinlichkeit'') is ``the probability of the macrostate;'' cf. \cite{GL}.
 The reference to ``probability,'' if understood in terms of ``relative frequency'' of occurrence in 
independent and identically distributed trials, is somewhat problematic
in a setting where the $N$ body system is all the matter in the one and only universe. 
 It is more appropriate to refer to $W$ as a \emph{typicality measure} for the macrostate: simply the size of the region 
in $N$ particle phase space consisting of microstates which all give the same macrostate under consideration.
 If one were to consider, say,  a simple fluid in local thermal equilibrium \cite{GGL}, a macrostate at time $t$ 
would be the collection $(\nu(\sV,t), \eps(\sV,t), \uV(\sV,t))$ consisting of the particle density ($\nu$), 
energy density ($\eps$), and velocity field ($\uV$) of the fluid. 
 If the system is a dilute gas, possibly not in local thermal equilibrium, then its (kinetic) macrostate at time $t$ is 
given by $Nf(\sV,\pV,t)$ (sometimes referred to as a ``mesoscopic state''). 

 Boltzmann also had the important insight that for a macroscopic system of $N\gg 1$ particles not at a phase transition the 
measure of the $N$ particle phase space of microstates which correspond to the thermal equilibrium state is essentially the
full size of the available region in $N$ body phase space. 
 Ignoring angular momentum conservation for simplicity, in our setting 
this phase space region is the hypersurface $\{H^{(N)} = E\}$ in $T^*({\mathbb S}_R^3)^N$,
where $E$ is the energy of the universe and where
\begin{equation}\label{Ham}
H^{(N)}  
= \sum_{1\leq k\leq  N} \frac{|\pV_k|^2}{2m} +
 \sum\sum_{\hspace{-0.7truecm}1\leq k < l\leq N} U(\chord{\, \qV_k - \qV_l\,})
\end{equation}
is the Hamilton function. 
 Right now $H^{(N)}$ is the sum of Newtonian kinetic energies and bounded continuous
pair interaction energies $U(\chord{\, \qV_k - \qV_l\,})$, 
expressed in terms of the phase space variables $\qV_k$ and $\pV_k$, $k\in\{1,...,N\}$.
 Recall that particle positions in ${\mathbb S}_R^3$ have been identified with vectors $\qV_k\in{\mathbb R}^4$ 
of Euclidean length $|\qV_k| = R$. 
 The Euclidean distance $\chord{\, \qV_k - \qV_l\,}$ is called their \emph{chordal distance}
between points $\qV_k$, $\qV_l$ $\in{\mathbb S}^3_R$; cf. Appendix A.

 To compute (in principle at least) the measure of a hypersurface in a high dimensional space, we 
recall that in school we learn that the volume of a three-dimensional ball of radius $r$ in
Euclidean space is $V(r)=\frac43\pi r^3$, and its surface area is $A(r)=4\pi r^2$. 
 Next recalling our college calculus courses, we notice that $A(r) = V^\prime(r)$, where the ${}^\prime$ means derivative. 
 In a similar vein, if $\boldsymbol{1}_X$ denotes the so-called \emph{indicator function} of the set $X\subset
T^*({\mathbb S}_R^3)^N$, which takes the value $1$ on $X$ and the value $0$ outside of $X$, then 
\begin{equation}\label{PHIofE}
\Phi(E) := 
\int_{T^*({\mathbb S}_R^3)^N}
\boldsymbol{1}_{\{H^{(N)} \leq E\}} {\textstyle\frac{\dd^3{p}_1...\dd^3{q}_N}{h^{3N}}}
\end{equation}
denotes the $N$ body phase space measure of the region $\{H^{(N)} \leq E\}$ in $T^*({\mathbb S}_R^3)^N\!$, 
normalized by $h^{3N}$, and so $\Phi^\prime(E)$ now yields the hypersurface measure of the hypersurface $\{H^{(N)} = E\}$. 

 To obtain Boltzmann's thermal equilibrium entropy (\ref{SlogW}) of such an $N$ body system having energy $E$, we now have
to take the logarithm of $\Phi^\prime(E)$ --- essentially, though not quite!
 Since $\Phi^\prime(E)$ is not a dimensionless quantity, we multiply it by a reference energy unit, say $mc^2$. 
 Also, (\ref{PHIofE}) overcounts the physically relevant phase space size by a factor $N!$ (physical particles do not carry
labels), which we divide out. 
 And so, for the thermal equilibrium state of a classical system of $N$ point particles in a spherical universe, 
Boltzmann's (\ref{SlogW}) (updated with Planck's $h$, $\kB$, and Gibbs' $N!$) essentially becomes 
\begin{equation}\label{SBergo}
\SB(E,N) 
= \kB \ln\Big[ {\tst\frac{1}{N!}}mc^2 \Phi^\prime(E) \Big]\, .
\end{equation}

%

 We next connect (\ref{SBergo}) with (\ref{SB}). 
 For non-singular pair interactions $U$ it was rigorously shown in \cite{KieRMP} that, if $N$ is large enough, with
energy scaling $E=N^2\veps$, where $\veps$ is a fixed parameter, and momenta rescaled as $\pV\mapsto\sqrt{N}\pV$,
then (with $o(N)$ meaning: $o(N)/N\to 0$ as $N\to\infty$) we have
\begin{equation}\label{maxS}
 \SB(N^2\veps,N) 
= 
\max_{f\in \Asp_{\veps}}\pzcS_{\mathrm{B}}(f;N)
+o(N)\, ,
\end{equation}
where $\Asp_{\veps}$ is the admissible set of normalized density functions $f(\sV,\pV)$ for which also $f\ln f$ 
is integrable and for which the energy of $f$, given by
\begin{equation}\label{eOFf}
\hspace{-10pt}
\pzcE(f) 
=\!\! \iint \tfrac{1}{2m}|\pV|^2 f(\sV,\pV)\dd^3p\dd^3\sV
+ \!\! \iiiint{\tfrhalf}
U(|\sV-\tilde{\sV}|)f(\sV,\pV)f(\tilde{\pV},\tilde{\sV})
	\dd^3p\dd^3s \dd^3\tilde{p}\dd^3\tilde{s},
\end{equation}
satisfies $\pzcE(f) = \veps$; here, each $\int\!\!\!\int$ means an integral over $T^*{\mathbb S}^3_R$, cf. (\ref{SB}).
 Note that formula (\ref{eOFf}) is the kinetic theory analog of (\ref{Ham}).

 Formula (\ref{maxS}), which links Boltzmann's entropy (\ref{SBergo}) of the statistical (thermal) equilibrium state of 
the spherical $N$ body universe with the entropy functional (\ref{SB})
(i.e. Boltzmann's $H$ function), explicates the celebrated 

\centerline{\sc{Maximum Entropy Principle}:}

\centerline{\emph{The thermal equilibrium state of an isolated system}}

\centerline{\emph{is a macrostate of highest possible Boltzmann entropy.}}

 We are now ready to inquire into Clausius' proposal 
that the fate of the universe is to end up in its highest possible entropy state. 
 We will ignore all but the purely Newtonian gravitational interactions,
treated as singular limit of bounded continuous pair interactions. 
 On $\Sset^3_R$ they read $U(\chord{\, \qV_k - \qV_l\,}) = - \frac {G m^2}{\chord{\, \qV_k - \qV_l\,}}$, 
which appears to be the same as in $\Rset^3$ except that the distance is the four-dimensional Euclidean 
distance of points in $\Sset^3_R\subset\Rset^4$ (Appendix A).

\section{The non-relativistic classical universe}\vspace{-5pt}

Consider a finite number $N$ of identical particles with Newtonian gravitational interactions 
on ${\mathbb S}^3_R$, having total energy $H=E$ at a fixed time $t$, where 
\begin{equation}\label{Ham3d}
H^{(N)}  = \sum_{1\leq k\leq  N} \frac{|\pV_k|^2}{2m} - 
 \sum\sum_{\hspace{-0.7truecm}1\leq k < l\leq N} \frac {G m^2}{\chord{\, \qV_k - \qV_l\,}}
\end{equation}
is the Hamilton function.
 Recall that points on ${\mathbb S}^3_R$ have been identified with vectors $\qV_k\in{\mathbb R}^4$ 
of Euclidean length $|\qV_k| = R$, and that $\chord{\, \qV_k - \qV_l\,}$ is the chordal distance function on $\Sset^3_R$.

 In the previous section we mentioned that when 
$- \frac {G m^2}{\chord{\, \qV_k - \qV_l\,}}$ is replaced by a continuous $U({\chord{\, \qV_k - \qV_l\,}})$,
the asymptotic large $N$ expansion of the Boltzmann entropy (\ref{maxS})--(\ref{eOFf}), with $\pzcS_{\mathrm{B}}(f;N)$ 
given by (\ref{SB}), has been established in \cite{KieRMP}; note that it does not seem solvable in closed form.
 In the following, when we simply write the Newtonian gravitational pair interaction, it has to be understood as singular
limit of a family of bounded continuous pair interactions.

 The expression (\ref{SBergo}) for Boltzmann's entropy of the statistical equilibrium state can be integrated in the
$\pV$ variables over $N$ copies of ${{\mathbb R}^{3}}$, which gives us 
\begin{equation}\label{SBergoINT}
\SB(E,N) 
=   \kB \ln \Big[C
\int_{({\mathbb S}^3_R)^N}\!\!
\Big(E +\!  \tst{\sum\limits_{ k < l}} \frac {G m^2}{\chord{\,\qV_k - \qV_l\,}}
\Big)_+^{\frac{3N -2}{2}}\!\!\dd^3{q}_1...\dd^3{q}_N  \Big]
\end{equation}
where $C= \frac{mc^2\sqrt{2\pi m}^{\,3N}}{N!\Gamma\left(\frac{3N}{2}\right)h^{3N}}$, and 
the subscript $_+$ means the positive part (i.e. negative values are replaced by 0).
 By direct inspection one sees that integral (\ref{SBergoINT}) exists only for $N\leq 2$, while it diverges for $N\geq 3$.
 Thus, in particular:
\vspace{5pt}

 \emph{The Boltzmann entropy of a finite classical universe of $N\gg 1$ Newtonian }

 \emph{point particles in $\Sset^3_R$ is unbounded above for any finite $E$.} 
\vspace{5pt}

 \emph{Conclusions}: 
 Clausius' first law of the universe 1) is still mathematically meaningful for this 3-dimensional classical toy universe, 
but his law 2) is not, because there is no maximum entropy state.
 Yet we can accomodate the spirit of Clausius' law 2) --- the increase of entropy --- by replacing it with
\medskip

\hspace{1truecm}{Law $2^\prime$: \textit{The entropy of the world increases beyond any bound.}}
\medskip

 In a three-dimensional spherical space $\Sset^3_R$, according to laws 1) and $2^\prime$) 
the fate of the classical toy universe is not, to reach a maximum entropy state after which 
all macroscopic evolution ceases forever (in the sense that the dynamics becomes static:
the famous ``heat death''), but a never-ending entropy-raising evolution ---
unless the entropy blows up to infinity in a finite amount of time, at which point the macroscopic evolution may 
cease in a different sense, having reached a ``singular state,'' which may be 
considered to be a different type of ``heat death.'' 

 While entropic considerations do not yield the time scales involved, they do offer insights into
the qualitative type of evolution. 
 The first insight in this direction came in 1962 in the celebrated paper \cite{antonov}. 
 Recall that (\ref{SBergoINT}) is the singular limit of a family of similar integrals in which $\frac{1}{\chord{\qV-\qV'}}$
is replaced by a regularized interaction.
 As we already know, for the regularized interactions the asymptotic (as $N\to\infty$) expansion of (\ref{SBergoINT}) 
is given by (\ref{maxS})--(\ref{eOFf}), with $\pzcS_{\mathrm{B}}(f;N)$ given by (\ref{SB}), and so it follows that the 
singular limit of this maximum Boltzmann entropy variational principle, namely to maximize (\ref{SB}) constrained with 
$\pzcE(f) =\veps$, where
\begin{equation}\label{eOFfG}
\hspace{-13pt}
\pzcE(f) 
=\!\frac{1}{2m} \iint\! |\pV|^2 f(\sV,\pV)\dd^3p\dd^3\sV
- \frac{Gm^2}{2} \!\! \iiiint
\frac{f(\sV,\pV)f(\tilde{\pV},\tilde{\sV})}{|\sV-\tilde{\sV}|}
	\dd^3p\dd^3s \dd^3\tilde{p}\dd^3\tilde{s},
\end{equation}
has no maximizing solution $f_\veps$.
 Antonov did not argue in this manner but proved directly that the Boltzmann entropy functional (\ref{SB}) 
constrained with $\pzcE(f) =\veps$, with $\pzcE(f)$ given in (\ref{eOFfG}), has no upper bound; 
in fact, he proved this for a gravitating ideal gas in a spherical container~\cite{antonov}, 
but his proof can be adapted to our model on $\Sset^3_R$. 

 We outline his strategy of proof.
 It is to break up the system into a small localized core, which collapses and whose 
gravitational energy becomes more and more negative, and a uniform halo which picks up that energy, thus
heating up and in the course of it boosting its entropy beyond any bound.
 Also the core gets hotter, but not as hot as the halo. 
 The curious thing about Antonov's proof is that the core loses mass while it shrinks, which is picked up by
the halo. 
 In this continuum approximation the mass of the core formally converges to zero when the entropy tends to $\infty$.

 Of course, in an $N$ body system in which each particle has mass $m$ there is a smallest possible mass of a ``collapsing core''
from which an infinite amount of gravitational energy could be extracted, namely a single pair of particles whose 
separation distance converges to zero. 
 A very detailed dynamical study of this scenario has been carried out in a monumental work by Heggie~\cite{heggie}. 
 Heggie's work indicates that an $N$-body system develops a tightly bound binary plus an expanded halo
containing all other particles. 
 The halo is heated at the expense of the binary, which gets bound together ever tighter through close
encounters with an occasional third particle.
 In the course of infinitely many such encounters the binary system shrinks to a single point, thereby
liberating an unlimited amount of gravitational binding energy. 
 The halo of the $N-2$ remaining point particles picks up the liberated energy in form of kinetic energy and
this carries the entropy to infinity.
 
  \emph{Final remark}: Even if the process which shrinks the binary to a point will take forever, in such a 
Newtonian universe matter becomes arbitrarily hot in the process and eventually distributed uniformly on $\Sset^3_R$.
 \newpage

\section{\hspace{-5pt}The non-relativistic quantum universe}\vspace{-5pt}

 The finding in the previous section of an unbounded entropy relies on a single pair of gravitating
point particles being able to orbit arbitrarily close to one another, which is allowed in
classical physics but not in quantum physics; just recall the hydrogen atom.
 And so one would expect that a system of $N$ gravitating point particles in quantum mechanics
should have an upper bound to their entropy, given by the quantum analog of Boltzmann's entropy (\ref{SBergoINT}), 
\begin{equation}\label{SBQtr}
\SB^{Q}(E,N) = \kB \ln {\rm Tr} P_{E,\triangle E};
\end{equation}
here, $P_{E,\triangle E}$ is the projector onto the subspace of Hilbert space spanned by energy eigenstates
with energy in a small interval $\triangle E$ centered on $E$.
(The size of $\triangle E$ does not matter as long as it contains very many eigenvalues. It can be 
chosen as small as one pleases (not zero) provided one makes $N$ large enough, correspondingly.)
 In fact there should be a maximum entropy state.

 So now consider non-relativistic quantum mechanics for fermions interacting via Newtonian gravity. 
 One may think of the particles as neutrinos, which have a tiny rest mass and spin 1/2 but no charge.
 The Hamiltonian is (\ref{Ham3d}), except that now $\pV_k = (\hbar/i)\boldsymbol{\nabla}_k$, 
where $\hbar$ is  Planck's constant $h$ divided by $2\pi$ and $\boldsymbol{\nabla}_k$ the gradient operator
in the $k$-th position variable. 
$H$ acts on antisymmetric $N$ particle wave functions (the Pauli exclusion principle for fermions). 
 By the same proof as given in~\cite{levyleblond} for $N$ such particles in $\Rset^3$,
one finds that also our Hamiltonian for $N$ particles in $\Sset^3_R$ is bounded below with
$\inf {\rm spec}\ H \propto -N^{7/3}$; see also~\cite{lieb}.
 Moreover, $H$ has a self-adjoint Friedrichs extension, and since $\Sset^3_R$ has finite volume,
$H$ has purely discrete spectrum for which Weyl's asymptotic law for the counting of eigenvalues holds. 
 As a consequence, for finite $N$ and $E>E_g$, where $E_g$ is the ground state energy, the
non-relativistic quantum mechanical analog (\ref{SBQtr}) of Boltzmann's entropy is finite. 

 The fermionic quantum analog of Boltzmann's entropy $\pzcS_{\mathrm{B}}(f;N)$ is  
\begin{equation}\label{SBQ}
\pzcS_{\mathrm{B}}^{Q}(f;N)
 = 2\pzcS_{\mathrm{B}}(f;N)
 -   2N \kB h^{-3}\! \int_{{\mathbb S}_R^3}\!\!\Big(\int_{T^*_{\sV}(\Sset^3_R)} \!
(1-h^{3}f) \ln (1-h^{3}f) \dd^3{p}\Big)\dd^3{s};
\end{equation}
the factor of 2 is due to the two spin states of each fermion. 
 Note that the density functions $f(\sV,\pV)$ now are restricted by the stabilizing bound $h^3 f <1$.

 The quantum analog of Boltzmann's maximum entropy principle 
(\ref{maxS})--(\ref{eOFf}) for gravitating fermions confined to a box $\subset\Rset^3$
has been rigorously derived in the 1970s by Walter Thirring and his school; see \cite{Thirring}, \cite{Messer}.
 Their analysis, adapted to our setting, yields the asymptotic expansion for the entropy (\ref{SBQtr}) in terms
of the quantum entropy of a monatomic, self-gravitating ideal Fermi gas on ${{\mathbb S}_R^3}$, viz.
\begin{equation}\label{maxSBQ}
 \SB^Q(N^{7/3}\veps,N) 
= 
\max_{f\in \Asp_{\veps}}\pzcS_{\mathrm{B}}^{Q}(f;N)
+o(N)\, ;
\end{equation}
here, $\Asp_{\veps}$ is the set of normalized $f(\sV,\pV)$ for which $f\ln f$ 
and $(1-h^3f)\ln (1-h^3f)$ are integrable and for which $\pzcE(f) = \veps$, with
$\pzcE(f)$ given in (\ref{eOFfG}).

 The Euler--Lagrange equations for this maximum quantum entropy principle yield for each $\sV\in\Sset^3_R$ that
$f(\sV,\pV)$ is a well-known Fermi--Dirac density on $T^*_\sV(\Sset^3_R)$ of an ideal Fermi gas. 
 The normalized particle density 
\begin{equation}
\rho(\sV) : =
\int_{{\mathbb R}^3}f(\sV,\pV)  \dd^3{p}\, 
\end{equation} 
in turn satisfies a nonlinear system of integral equations which, to the best of my knowledge, have not been 
discussed in the literature. 
 It is easy to see that they always admit the spatially uniform solution of the non-gravitating ideal Fermi gas,
with the difference that the gravitational interactions may shift the chemical potential by a contant amount.
 For sufficiently large energy $\veps$ their solution is unique --- hence the uniformly distributed 
ideal Fermi gas is the maximum entropy state when $\veps$ is large enough.
 It is also straightforward to show that when $R$ is large enough, then there is a special value $\veps=\veps_J$ 
at which the uniformly distributed ideal Fermi gas becomes linearly unstable to spatially non-uniform
disturbances --- this is precisely the analog of the Jeans criterion, see \cite{Jeans,KieAAM}.
 At this Jeans energy $\veps_J$ the  entropy maximizers will exhibit a second-order phase transition at which
an $\Sset^3$-parametrized family of $SO(3)$ invariant states bifurcates off of the spatially uniform perfect gas, 
breaking its $SO(4)[=SO(3)\times SO(3)]$ symmetry.
 One can also anticipate what happens at even lower energies, for sufficiently large $R$, by taking 
guidance from the detailed numerical studies in  \cite{StahlETal,Chavanis} 
of similar gravitating systems in a spherical container $\subset \Rset^3$.
 Namely, with decreasing $\veps$ the maximum entropy states become more and more concentrated in a continuous
manner, until a special value of  $\veps$ is reached at which two different $\Sset^3$-parametrized families of 
maximum entropy states exist.
 This is the point of a first-order phase transition in the merely $SO(3)$ invariant states
where the \emph{state} of maximum entropy changes discontinuously as function of $\veps$:
 fixing the parameter in $\Sset^3$, the maximum entropy state in the family connected to the Jeans bifurcation 
is moderately condensed and has the lower temperature, the other maximum entropy state has a strongly condensed core and
a dilute halo, and is much hotter.
	The first-order transition will be associated with local entropy maximizers in its neighborhood,
which can be interpreted as metastable states. 
	These will terminate at their respective spinodal points, which for the low
temperature states is determined by the analog of the Emden-Jeans criterion \cite{Emden}. 
 Continuing to lower energies yet the core-halo state carries the highest entropy and eventually 
condenses onto a white dwarf type ground state of ``monumental size,'' containing all matter of the universe.
\begin{figure}[H]
\centering
\includegraphics[scale=0.5]  
{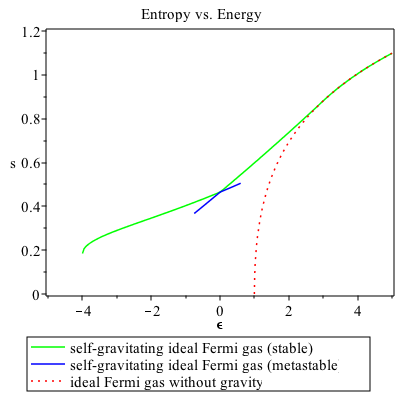}
\label{SvsEin3d}
\end{figure}

 Fig.~2 displays a qualitative sketch of the entropy $s$ as function of the energy $\veps$, 
of various states of the self-gravitating ideal Fermi gas on $\Sset^3_R$ with $R$ much larger than the
particles' Compton length.
 The maximum entropy graph is shown in green.
 Also shown (in blue) is the entropy for the meta-stable states which are merely local entropy maximizers,
and (in red, dotted) the maximum entropy of the ideal Fermi gas without gravity (i.e., $G=0$). 
 
 \emph{Conclusions}: 
 Clausius' two laws of the universe 1) and 2) are mathematically meaningful for this three-dimensional non-relativistic
quantum toy universe.
 They predict its fate to be a ``heat death'' when all macroscopic evolution ceases.
 Depending on the energy content of the world, this ``heat death'' can be the uniform distribution of matter 
in $\Sset^3_R$ (having $SO(4)$ invariance) with $\approx$ Maxwellian velocity distribution and a (hot) temperature --- 
this would happen at high energies.\footnote{Alternatively, the uniform matter state also maximizes entropy when $R$ is
  very small. 
 This raises the question whether the most typical (i.e. representative) macrostate of the early universe was perhaps
 uniform --- eliminating any need for an inflationary phase to explain the essential uniformity of the cosmic microwave 
background radiation!}
 Or it can be a non-uniform, but small gradients distribution of matter in $\Sset^3_R$ (having $SO(3)$ invariance),
with less Maxwellian and more Fermi--Dirac like velocity distribution and moderate temperatures --- this would happen 
at intermediate energies. 
 Or, it can be a very strongly condensed core distribution of matter in $\Sset^3_R$ surrounded by a very diluted halo
(both parts still having $SO(3)$ invariance);
the core will have a recognizable Fermi-Dirac velocity distribution, and low temperature --- this would happen 
at very low energies. 

 At the first order phase transition point, the description is more complicated. 
 In particular, the state with a highly condensed core and very dilute atmosphere
is hotter than the less condensed, small gradients state.

\section{\hspace{-7pt}The ``special-relativistic'' quantum universe}\vspace{-5pt}

 In this final section before coming to general relativity, 
inspired by Chandrasekhar's theory of white dwarf stars we take special relativity 
into account while still working with Newtonian gravity.
 Yet, instead of using a many-body Dirac equation we work with the ``pseudo-relativistic'' Hamiltonian
\begin{equation}\label{HamSQRrtKG}
H^{(N)}  = \tst \sum\limits_{1\leq k \leq N} \sqrt{m^2c^4 - \hbar^2c^2\Delta_k}
- {G m^2} \sum\!\!\!\sum\limits_{\hspace{-.5truecm}1\leq k < l\leq N} 
\tst\frac{1}{\chord{\, \qV_k - \qV_l\,}},
\end{equation}
acting on anti-symmetric wave functions for $N$ fermions in $\Rset^3$.
 A rigorous analysis for the ground state of (\ref{HamSQRrtKG}) has been carried out 
by Lieb and Thirring~\cite{liebthirring}.
 The Hamiltonian (\ref{HamSQRrtKG}) does have a ground state only if $N<N_{\mathrm{{Ch}}}$, with
\begin{equation}
N_{\mathrm{{Ch}}} = C\frac{9}{16\sqrt{\pi}}
\Big( \frac{h c}{Gm^2}\Big)^{3/2} 
\end{equation}
where $C$ is determined by solving a nonlinear PDE numerically; $C = 0.87..$ if we neglect spin state counting.
 If $m$ is the neutron mass, then $N_{\mathrm{{Ch}}}\approx 10^{58}$, which essentially yields Chandrasekhar's
maximum mass of a white dwarf star. 

 What does this imply for the entropy (\ref{SBQtr})? 

 For $N<N_{\mathrm{{Ch}}}$ the analogous reasoning as in the non-relativistic case leads again to a finite quantum
Boltzmann entropy (\ref{SBQtr}) of the statistical equilibrium state.
 However, this argument fails in the supercritical case $N>N_{\mathrm{{Ch}}}$ relevant to our universe.
 Indeed, as argued in \cite{KieISCHIA}, for $N > 1.2N_{\mathrm{{Ch}}}$ and any given $E$ 
the quantum Boltzmann entropy (\ref{SBQtr}) diverges; the factor $1.2$ is not optimal. 
 We do not repeat the full argument here but note some important points. 

 The argument that the quantum Boltzmann entropy (\ref{SBQtr}) diverges
for $N > 1.2N_{\mathrm{{Ch}}}$ in the Hamiltonian (\ref{HamSQRrtKG}) is based on a rigorous proof 
that the quantum analog of Boltzmann's maximum entropy principle for the semi-classical continuum approximation 
to this $N$ fermion system has no upper bound. 
 The strategy of proof is adapted from Antonov's~\cite{antonov}, see section 4.
 Note that the existence of Chandrasekhar's limit mass, viz. $N_{\mathrm{{Ch}}}$, makes the proof more tricky 
than for the classical non-relativistic continuum approximation, where no such limit mass exists. 
 One now takes $N>1.2N_{\mathrm{{Ch}}}$ fermions and split them into two subsystems, one with $N_1>N_{\mathrm{{Ch}}}$ 
particles and one with $N_2$ particles, such that $N_1 +N_2 = N$.
 While $N_2$ need not exceed $N_{\mathrm{{Ch}}}$, it cannot be arbitrarily small for a semi-classical approximation
to hold, but $N_2 > 10^{-20}N_{\mathrm{{Ch}}}$ will do.
 Both subsystems are placed far apart so that their mutual gravitational energy can be ignored in the argument.
 Then one can find semi-classical continuum approximations $f_1$ and $f_2$ to the fermion densities in the two
subsystems which obey the exclusion principle $h^3 f_n <1$ for $n = 1,2$, such that the total quantum Boltzmann 
entropy of $f_1$ and $f_2$, given by $\pzcSB^Q(f_1,N_1) + \pzcSB^Q(f_2,N_2)$, 
surpasses any prescribed value $S$ while obeying the energy constraint 
$\pzcE^{SR}(f_1,N_1) + \pzcE^{SR}(f_2,N_2) = E$, where now 
\begin{alignat}{1}
\pzcE^{SR}(f;N)
 = \ & N^{\frac{4}{3}} 
\iint\sqrt{m^2c^4 + p^2c^2}\, f(\sV,\pV)  \dd^3{p}\dd^3{s} \nonumber \\
& - N^2  \frac{Gm^2}{2} \iiiint
\frac{f(\sV,\pV)f(\tilde{\pV},\tilde{\sV})}{|\sV-\tilde{\sV}|}
	\dd^3p\dd^3s \dd^3\tilde{p}\dd^3\tilde{s}
\end{alignat}
is the (pseudo-) special-relativistic energy functional of $f$;
note that here we have retained $N$ --- the existence of $N_{\mathrm{{Ch}}}$ makes it clear that a simple
rescaling of space and energy scales cannot be used to absorb $N$ into a scaling factor as done in the non-relativistic
models.
 The extra $N^{1/3}$ factor at the kinetic energy is all that remains from Pauli's exclusion principle for $N$ fermions, 
which at the microscopic level rules out that two or more fermions are in the same single-particle  state.
 In this sequence, subsystem 1 is a massive core which collapses, not losing any mass in the process, but
liberating gravitational energy, which is picked up by a halo that does not
collapse, heating up instead and in the course boosting its entropy beyond any bound. 

 \emph{Conclusions}: 
 Clausius' first law of the universe 1) is again mathematically meaningful for this 3-dimensional quantum toy universe
of $N\gg 1$ particles, but his law 2) holds only if $N< N_{\mathrm{{Ch}}}$.
 Since there is no maximum entropy state for $N>1.2N_{\mathrm{{Ch}}}$ fermions, once
again one has to accomodate the entropy increase in the spirit of Clausius by replacing law 2) with the law $2^\prime$) 
stated in section 4.

 This (pseudo-) special-relativistic three-dimensional quantum toy universe resembles its non-relativistic
version only if the universe contains not many more particles than our sun (if $m$ is the neutron mass).
 But our galaxy alone contains about $10^{11}$ stars, and there are hundreds of billions of galaxies in the universe, and
for such huge $N$ our (pseudo-) special-relativistic three-dimensional quantum toy universe resembles the non-relativistic
three-dimensional classical toy universe of section 4 more closely.
 The main difference is that in the classical setting a single pair of particles converging onto the same location
suffices to boost the entropy beyond any bound, while in the special-relativistic 
quantum setting one needs more than $N_{\mathrm{{Ch}}}$ fermions.

 \emph{Additional remarks}: For the regime $N< N_{\mathrm{{Ch}}}$ the semi-classical continuum approximation to the
ground state has been rigorously justified by Lieb and Yau~\cite{liebyau} who derived Chandrasekhar's 
structure equations of a white dwarf star from an analysis of the ground state of (\ref{HamSQRrtKG})
in a suitable large-$N$ continuum limit; see also~\cite{lieb}. \vspace{-10pt}
\newpage

        \section{On the significance of the black hole entropy}
        \label{sec:cosmology}

 So far we have exclusively considered Newtonian gravity of $N$ point particles in a static spacetime $\Sset^3_R\times \Rset$, 
and only the dynamics of the particles became more and more realistic
from section to section: we began with non-relativistic Newtonian mechanics;
next we used, first non- and then (pseudo-) special-relativistic quantum mechanics.
 However, Newton's theory misses the important contribution of black holes to classical gravity theory,
discovered in general relativity.

 The first nontrivial exact solution to Einstein's field equations discovered, the 
Schwarzschild solution, is a spacetime containing a black hole (and a white hole, and which has two spacelike
separated asymptotically flat (Minkowskian) regions).
 In Schwarzschild coordinates $(t,r,\theta,\phi)$ the asymptotically flat region $\{r> 2GM/c^2\}$ has metric~\cite{pauli,MTW}
\begin{equation}\label{BH}
ds^2 = - \Big(1 - \frac{2GM}{c^2r}\Big)c^2dt^2 
+\Big(1 - \frac{2GM}{c^2r}\Big)^{-1}dr^2 + r^2\big(d\theta^2 + \sin^2\theta d\phi^2\big) ,
\end{equation}
where $M$ is the mass ``as seen from far away'' from the hole, i.e. when $r\gg 2GM/c^2$.
 The metric (\ref{BH}) is singular at $r = 2GM/c^2$, but this is an artifact of the coordinates used. 
The spacetime itself is regular at the three-dimensional hypersurface $\{r = 2GM/c^2\}$, which 
defines the event horizon of the black hole. 
 What makes the black hole ``black'' is the feature that no lightlike geodesic which starts at a spacetime event
inside the event horizon will be able to reach what relativists call ``future null infinity'' of the spacetime, while
for every event outside the event horizon, there exists a lightlike geodesic which does.
 In more colloquial terms, once inside a black hole, you can't communicate with the world outside of it using electromagnetic
radiation.

 This raises an important question: ``Even though it cannot send electromagnetic
signals to the outside world, can matter which falls into a black hole
transfer energy to, and thereby raise the entropy of, matter outside of it?''

 Roger Penrose has shown how a certain amount of energy could be extracted, in principle, from a rotating black hole. 
 So at least in the most general sense of the first important question, the answer is ``Yes.''
 
 A curious aspect of the Schwarzschild metric is the following.

 A $t=const.$ section of its event horizon is a two-dimensional sphere with area $4\pi (2GM/c^2)^2$, 
and the famous \emph{Schwarzschild radius}
\begin{equation}
R_{\mathrm{S}} : =  \frac{2GM}{c^2}\, 
\end{equation}
is the Euclidean radius of a sphere $\Sset^2_{R_{\mathrm{S}}}\subset\Rset^3$ with the same area.
 Any sphere $\{(t,r,\theta,\phi): t=const.\, \&\, r=  R_{\mathrm{S}}\}$ is a so-called \emph{marginally trapped surface}. 
 It is common in the astrophysics community to think of black holes as evolutionary three-dimensional 
``quasi-objects'' hidden inside a marginally trapped two-dimensional surface, and the folklore says
that two such black holes can merge, but no black hole can split into two.
 In this vein, if one compares, a spacetime which contains two distant Schwarzschild-type black holes 
of masses $M_1$ and $M_2$, with a spacetime containing a single such hole whose mass is $M = M_1 + M_2$, then 
the surface area of a $t=const.$ section of its event horizon is
\begin{equation}
A = 4\pi R_{\mathrm{S}}^2 = 16\pi \frac{G^2}{c^4}(M_1+M_2)^2,
\end{equation}
and by the simple inequality $(M_1+M_2)^2 > M_1^2 + M_2^2$ (both $M_k>0$) we have $A > A_1 + A_2$.
 Accepting the folklore then we have in front of us
the black hole analogues of the first and second laws of thermodynamics, the so-called first and 
second laws of black hole thermodynamics: black hole mass plays the role of energy, the area of the marginally
trapped surface plays the role of entropy.\footnote{Einstein's $E=mc^2$ makes it plain that mass must
  play the role energy, but entropy as a surface quantity is a novelty.}
 This is the gist of it; for rotating black holes there are some modifications \cite{heusler}.

 This purely formal analogy to conventional thermodynamics has led Bekenstein~\cite{bekensteinA} to suggest 
that a black hole of mass $M$ actually \emph{has} an entropy $\propto M^2$, which Hawking~\cite{hawking} computed 
to be\footnote{In the pertinent formula (1) of \cite{KieISCHIA} a factor $8\pi^2$ is missing.}
\begin{equation}\label{BHS}
 S_{\mathrm{BH}} = \kB\frac{4\pi G}{\hbar c} M^2. 
\end{equation}
 To arrive at (\ref{BHS}) Hawking went beyond classical general relativity, but it is fair to say that his heuristic
derivation  still awaits a rigorous foundation.

 Yet accepting the Bekenstein--Hawking entropy as a physical entropy, the second important question one will have to 
address is: ``Which role does black hole entropy play in the assessment of Clausius' law 2)?''

 In his book ``The emperor's new mind''~\cite{penroseR} Roger Penrose argues (p.338)
that when atomistic matter starts uniformly distributed
over a spherical space $\Sset^3_R$, it will develop clumps under its gravity, and those clumps will coalesce, and 
the universe will evolve in the spirit of Clausius' law 2) into a state of highest entropy --- which according to
Penrose will be a black hole that has swallowed up all the matter of that universe, 
having a Bekenstein--Hawking entropy (\ref{BHS}) with $M$ the mass of the universe.
 Taking $M=10^{80}m$ with $m$ the mass of a proton, Penrose computes (pp.342/3) the maximum possible entropy 
of our universe to be $10^{123}\kB$.  
 In Penrose's scenario the two laws of Clausius, 1) and 2), are assumed valid, resulting in the ``heat death'' of the universe,
although this ``heat death'' has little in common with what Clausius and his contemporaries had envisioned. 

 Incidentally, like Boltzmann ignoring Poincar\'e's recurrence time as irrelevant to the explanation of the fate of our material 
universe, Penrose points out that the evaporation of a black hole by Hawking radiation is irrelevant to his argument 
because of the stupendously large time scales involved. 

 Astoundingly, while the entropy of a universe filled with $N$ gravitating fermions which have a special-relativistic 
kinetic energy but interact with Newtonian gravity is unbounded above (for $N> 1.2 N_{\mathrm{{Ch}}}$), its 
more realistic general relativistic version seems to have a bounded entropy!
 However, if we would assume that the fate of the universe in the semi-classical (pseudo-)special-relativistic theory 
is a collapse of all the matter onto itself (the closest analog in that model to a universe whose matter has formed a single 
black hole), we would get an upper bound on its entropy \cite{KieISCHIA}, viz. 
\begin{equation}
\SB^Q(f;N) \leq  \kB 2N \ln N + O(N)\, .
\end{equation}
 This raises the question whether in general relativity one might also get higher entropies by splitting the system 
into a collapsing core and a halo which receives the liberated energy from the core and carries the entropy to infinity. 
 We need to inspect the Bekenstein--Hawking entropy more closely.

 Bekenstein's reasoning for his black-hole entropy formula is based on the proposition that physics is concerned only 
with the world outside the event horizons of all the black holes in the universe, as most physicists seem to have 
argued it would --- back then. 
 But then, whenever a black hole swallows a piece of matter, it also swallows with it its entropy, in the process of which 
the entropy of the matter outside the event horizons decreases --- violating the second law even in its weakest form: 
\vspace{3pt}

\hspace{1truecm}{$2^{\prime\prime}$) \textit{The world evolves such that its entropy does not decrease.}}
\vspace{10pt}

\noindent
To rescue $2^{\prime\prime}$) 
Bekenstein~\cite{bekensteinA} proposed that the ``entropy of the universe'' (at any instant of some cosmic time) 
is the sum, of the entropy of the ``matter'' outside the event horizons of the black holes, and of the entropy 
of those black holes. 
 If this sum obeys the second law at least in its form $2^{\prime\prime}$), insisting that this is a law of the universe, 
then it is logically conceivable that the hole will gobble up all ``matter,''  leaving only the black hole entropy;
hence Penrose's estimate for the maximum entropy of this $\Sset^3_R$ based universe model. 

 However, the Bekenstein-Hawking entropy of a black hole which has just swallowed a piece of ``matter'' is
not always larger than, or even equal to, the entropy of the orginal black hole plus the entropy
of the piece of ``matter'' before it was swallowed.
 This is readily demonstrated by considering a variation of the reasoning where one compares 
the areas of the marginally trapped surfaces 
of two Schwarzschild-type black holes with the surface area of the marginally trapped surface resulting after merger.
 Namely we now compare the sum of the Bekenstein--Hawking entropy of a Schwarzschild-type black hole of mass $M$ and the
entropy of a black body radiation occupying a stove of volume $V$ with the black hole entropy after this chunk of black body 
radiation has been swallowed up (we idealize the walls of the stove to have neglible mass and entropy; in fact, one doesn't need
a ``stove'' because the universe is already filled with a black body radiation.)
 The entropy $S_{bb}$ of such a photon gas expressed in terms of its energy $U_{bb} =: M_{bb}c^2$ is
\begin{equation}\label{bbS}
 S_{bb} = \kB \frac43\left(\frac{\pi^2 Vc^3}{15\hbar^3}\right)^{\frac14} M_{bb}^\frac34. 
\end{equation}
 And so, writing $S_{bb}(M_{bb}) := C_{bb}M_{bb}^\frac34$ and also $S_{\mathrm{BH}}(M) := C_{\mathrm{BH}} M^2$, we find 
\begin{equation}\label{masterIDENTITY}
C_{\mathrm{BH}} (M+M_{bb})^2 -  C_{\mathrm{BH}} M^2 - C_{bb}M_{bb}^\frac34 = 
C_{\mathrm{BH}} (2 MM_{bb} + M_{bb}^2) - C_{bb}M_{bb}^\frac34, 
\end{equation}
which makes it plain that
\begin{equation}\label{secondLAWfails}
S_{\mathrm{BH}}(M+M_{bb}) < S_{\mathrm{BH}}(M) + S_{bb}(M_{bb})  \quad \mbox{\textit{if}}\  M_{bb}\ 
\mbox{\textit{is\ sufficiently\ small}}!
\end{equation}
  Using the current temperature $T_{\mbox{\tiny{CMB}}}$ of the cosmic microwave background radiation, 
the criterion (\ref{secondLAWfails}) can be rephrased thus: 
\emph{if the black hole mass $M < \frac{\hbar c^3}{6\pi G \kB T_{\mbox{\tiny{CMB}}}}\approx 6\times 10^{22}{\mathrm{kg}}$,
then the BH entropy decreases by swallowing some CMB radiation.} 
 The borderline mass is about that of our moon.

  Curiously, Bekenstein in \cite{bekensteinA} already came to an equivalent conclusion, but then argued that 
statistical quantum fluctuations would invalidate the conclusion. 
 This may very well be so, but given the preliminary state of any investigation into the realm of quantum physics
in black hole spacetimes it seems fair to say that the jury seems still out on this case.

\medskip

 \emph{Conclusions}: 
  There are several important conclusions to be drawn from our discussion, 
about the evolution of such a universe model on the time scale of its existence, from the ``big bang'' until the ``big crunch,'' 
which certainly are much shorter than any ``Poincar\'e recurrence time'' or such.

 First of all, assuming law $2^{\prime\prime}$) 
holds in a general-relativistic universe model which describes the evolution of matter 
that was uniformly distributed over $\Sset^3_R$ initially, with entropy understood as the sum of the
entropy of the matter outside of the event horizons of all the black holes, plus the entropy of the black holes,
it is not yet clear whether the stronger Clausius' law 2) or its weakened version $2^{\prime}$) hold, too.
 What is clear, though, by letting $M+M_{bb}$ in  (\ref{secondLAWfails}) be the total mass in this universe model,
is the following:

\hspace{1truecm}\emph{A black hole which contains all ``matter'' of the universe}

\hspace{1truecm}\emph{is not the maximum entropy state of  such a universe!}

 Second, it is also conceivable that the fate of any ``matter'' initially outside of the event horizon of a
black hole in this closed universe model is eventually to end up inside a single black hole. 
 It may well turn out that the law $2^{\prime\prime}$) does not hold if one defines physics to 
be concerned only with what's outside of an event horizon, but $2^{\prime\prime}$), and possibly $2^\prime$) 
or even 2), may well hold (on the stipulated time scales) if we do not ignore the entropy of ``matter'' inside 
the event horizons of the black holes!
 Why should physics end at the event horizon?
 General relativity allows us to inquire into the fate of matter which has crossed an event horizon. 
 True, according to general relativity we will not be able to have a space probe explore this fate in situ and 
have its findings sent to us who reside outside of the event horizon. 
 But to insist that each and every logically coherent consequence of a physical theory,
for it to be acceptable as physics has to be ``directly measurable'' 
and ``communicable to wherever we are,'' seems to me too narrow a definition of what physics is about.
\vspace{-10pt}

\vfill
\noindent
{\textbf{Acknowledgment}:} 
 I thank Valia Allori for the invitation to write this article.
 It is based partly on \cite{KieISCHIA}, supported by NSF grant DMS-9623220, and
partly on \cite{KieJSTAT}, supported by grant DMS-0807705, updated by findings since. 
 The opinions of the author expressed in this paper are not necessarily those of the NSF. 
 I am indebted to Shelly Goldstein and Joel Lebowitz for teaching me Boltzmann's insights into the 
inner workings of the universe.
 I also thank Friedrich Hehl for reference \cite{heusler}.
 Lastly, I thank two anonymous referees for their helpful comments.
\newpage


\appendix
\section*{Appendices}
\addcontentsline{toc}{section}{Appendices}
\renewcommand{\thesubsection}{\Alph{subsection}}
\numberwithin{equation}{subsection}
\subsection{The Laplace operator on $d$-dimensional spheres}
 In $\Rset^3$, the Newtonian pair interaction energy $-Gm^2\frac{1}{|\hat{\sV} -\check{\sV}|}$ is, 
up to the factor $Gm^2$, the Green function for the Laplace operator $\Delta_{\Rset^3}$ on $\Rset^3$; i.e. 
\begin{equation}\label{greenDEF3dFLAT}
- \Delta_{\Rset^3} \frac{1}{|{\sV} -\check{\sV}|}
 = 4\pi \delta_{\{\check{\sV}\}}^{(3)}({\sV}) 
\end{equation}
in the sense of measures. 
	Here, ${|\sV -\check{\sV}|}$ denotes Euclidean distance in $\Rset^3$, and 
$\delta_{ \{\check{\sV}\}}^{(3)}(\sV)$ denotes the Dirac point measure at $\check{\sV}\in\Rset^3$, which means that for
any open Lebesgue set $\Lambda\subset \Rset^3$ we have
\begin{equation}\label{deltaDEF}
\int_\Lambda \delta_{\{\check{\sV}\}}(\sV)\dd^3s 
= 
\left\{ \begin{array}{rl} 1 & {\rm if}\ \check{s}\in \Lambda \\
			  0 & {\rm if}\ \check{s}\not\in \Lambda  
	\end{array}
\right. 
\end{equation}

 Similarly, $-\frac{Gm^2}{R}\ln \frac{R}{|\hat{\sV}-\check{\sV}|}$, with ${|\sV -\check{\sV}|}$ again Euclidean distance in $\Rset^3$, 
is the Green function for the Laplace-Beltrami operator $\Delta_{\Sset^2_R}$ on the sphere $\Sset^2_R$, i.e.
\begin{equation}\label{greenDEF2dROUND}
-\Delta_{\Sset^2_R} \ln\frac{R}{|{\sV} -\check{\sV}|} = 2\pi \Big(\delta_{\{\check{\sV}\}}^{(2)}({\sV})-{\tst{\frac{1}{4\pi R^2}}}\Big)
\end{equation}
in the sense of measures. 
 The additive constant at r.h.s.\Ref{greenDEF2dROUND} is inevitable, due to the 
topology of $\Sset^2$ --- note that $\int_{\Sset^2}\Delta_{\Sset^2} \ln\frac{1}{|{\sV} -\check{\sV}|} \dd^2s = 0$,
and that $\int_{\Sset^2_R} \dd^2s = 4\pi R^2$.

 Physicists would be tempted to interpret the term $- {\tst{\frac{1}{4\pi R^2}}}$ at r.h.s.\Ref{greenDEF2dROUND} as
a ``negative background mass density'' per particle which permeates ``space'' $\Sset^2_R$, but mathematically
it's just encoding the constant positive Gauss curvature of standard $\Sset^2_R$.

 Analogously, $-\frac{1}{|\hat{\sV} -\check{\sV}|}$, with ${|\sV -\check{\sV}|}$ now Euclidean distance in $\Rset^4$, 
is the Green function for the Laplace-Beltrami operator $\Delta_{\Sset^3_R}$ on the sphere $\Sset^3_R$; i.e.,
\begin{equation}\label{greenDEF3d}
- \Delta_{\Sset^3_R} \frac{1}{|\sV-\check{\sV}|} = 4\pi \Big(\delta_{\{\check{\sV}\}}^{(3)}({\sV}) - {\tst{\frac{1}{2\pi^2 R^3}}}\Big)
\end{equation}
in the sense of measures; note that $\int_{\Sset^3_R} \dd^3s = 2\pi^2 R^3$.

 To verify the Green function formulas for the spheres, note that for unit spheres,
${|\sV -\check{\sV}|}^2 = 2-2\cos\psi= 4\sin^2\frac{\psi}{2}$, where $\psi$ is the angle between $\sV$ and $\check\sV$;
now use the spherical angle representations for $\Delta_{\Sset^d}$.
\newpage

\subsection{The classical entropy of $N$ Newtonian particles in $\Sset^2_R$}\vspace{-5pt}

 The understanding of complicated issues is often aided by exactly solvable toy models --- caricatures 
of the real problem, yet clearly recognizable as such!
 In this vein it may be useful to record that the classical maximum entropy state can be computed 
exactly for a hypothetical world in which the dimension of physical space is two instead of three;
for the flat space analog, see \cite{Aly}. 
 The Newtonian gravitational interaction between a pair of particles with positions
$\qV_k$ and $\qV_l$ in $\Sset^2_R$ reads (Appendix A)
\begin{equation}\label{logG}
U(\chord{\, \qV_k - \qV_l\,}) = U_0 - \frac {G m^2}{R}\ln {\frac{R}{\chord{\, \qV_k - \qV_l\,}}},
\end{equation}
where $\chord{\, \qV_k - \qV_l\,}$ is now the chordal distance on $\Sset^2_R$.
 The constant $U_0$ is chosen for convenience so that $\int_{\Sset^3_R}U(\chord{\,\qV - \qV'\,})\dd^2q =0$ 
for each $\qV'\in\Sset^2_R$.

 Following verbatim \cite{KieJSTAT}, the analog of the maximum entropy principle
(\ref{maxS})--(\ref{eOFf}), with $\pzcS_{\mathrm{B}}(f;N)$ given by (\ref{SB}) with $\Sset^3_R$ replaced by $\Sset^2_R$
and $h^3$ by $h^2$, can (a) be rigorously derived from (\ref{SBergo}), (\ref{Ham}), with $U$ given by (\ref{logG}),
without any regularization, and (b) be evaluated completely, as follows:
 
	Any maximizer $f_{\veps}$ of $\pzcSB(f;N)$ over the admissible set $\Asp_{\veps}$ is of the form
\begin{equation}\label{fMaxwellBoltzmannMCerg}
f_{\veps}(\sV,\pV) 
= 
\sigma_{\veps}(\pV|\sV)\rho_{\veps}(\sV),
\end{equation}
with $\rho_{\veps}(\sV)$ solving the Euler--Lagrange equation
\begin{equation}
\label{fixPOINTEQrhoERGO}
\rho(\sV)
=
\frac{	\exp\left(-
	\int U(|\sV-\tilde{\sV}|)\rho(\tilde{\sV})\dd^2\tilde{s}/
\kB\vartheta_{\veps,\lambda}(\rho)
		\right)}
   {\int\exp\left(-
	\int U(|\hat{\sV}-\tilde{\sV}|)\rho(\tilde{\sV})\dd^2\tilde{s}/
        \kB\vartheta_{\veps,\lambda}(\rho) \right)\dd^2\hat{s}},
\end{equation}
where 
\begin{equation}
\label{ThetaEasFNCTLofRHO}
\kB\vartheta_{\veps}(\rho)
=
\veps 
-
{\tfrhalf} \iint U(|\sV-\tilde{\sV}|)\rho(\sV)\rho(\tilde{\sV})\dd^2{s} \dd^2\tilde{s}
\end{equation}	
is ($\kB\times$) the strictly positive ``temperature of $\rho$;'' in (\ref{fixPOINTEQrhoERGO}) and (\ref{ThetaEasFNCTLofRHO})
all integrals are over $\Sset^2_R$.
	The function $\sigma_{\veps}(\pV|\sV)$ is a scalar on $T^*_{\sV}\Sset^2_R$ and given by
\begin{equation}
\label{sigmaOFrhoERGO}
\sigma_\veps(\pV|\sV)
=
\left[{\tst{2\pi m\kB\vartheta_{\veps}(\rho_\veps)}}\right]^{-1}
\exp\bigl(-|\pV|^2/2m\kB\vartheta_{\veps}(\rho_\veps)\bigr).
\end{equation}	

 The Euler--Lagrange equation itself has many solutions, but two families of solutions can be 
explicitly stated in terms of elementary functions. 
 Happily these two families contain all the maximizers of $\pzcSB(f;N)$ over $\Asp_{\veps}$:
\begin{equation}
\label{LnulTYPf}
\hspace{-10pt}
f_{\veps}(\sV,\pV) 
= 
\left\{\!\! \begin{array}{ll} 
{\tst{\frac{1}{8\pi^2m\veps}}} \exp\bigl(- {\tst{\frac{1}{2m\veps}}}|\pV|^2\bigr)R^{-2}
\phantom{\Big(\Big.}
	& \!\!\!{;}\ \veps \geq \tfrquart\gamma
\\
{\tst{\frac{1}{2\pi^2m\gamma}}} \exp\bigl(- {\tst{\frac{2}{m\gamma}}}|\pV|^2\bigr)
	\left(R\cosh\zeta(\veps) - \aV\cdot\sV\sinh\zeta(\veps)\right)^{-2}
	 &\!\!\! {;}\ \veps < \tfrquart\gamma
	\end{array}
\right. 
\end{equation}
where we have set $\gamma := Gm^2/R$, and where 
$\aV\in\Sset^2\subset\Rset^3$ spans the arbitrary axis of rotational symmetry of 
$\rho_{\veps}(\sV)$, while $\zeta(\veps)>0$ is the unique positive solution of the fixed point equation
\begin{equation}
\label{zetaFIXptEQ}
\zeta = \left({\tst\frac{3}{2}- \frac{2\veps}{\gamma}}\right)\tanh\zeta.
\end{equation}

	Thus, for $\veps \geq \gamma/4$ the entropy maximizer is unique and identical with the spatially (on $\Sset^2_R$)
uniform thermal equilibrium of the non-gravitational ideal gas, having a Maxwellian momentum distribution at each 
$\sV\in\Sset^2_R$.
       For $\veps < \gamma/4$ the entropy maximizer is not unique and given by any one of the rotation-invariant
(about $\aV\in\Sset^2$) members of this two-parameter family.
	All these states have the same temperature, $\kB\vartheta=\frac14\gamma$, so their heat capacity is infinite. 

	Introducing the dimensionless energy $\eps:=4\veps/\gamma$, and writing for the entropy 
$\kB^{-1}\pzcSB(f_\veps,N)= - N\ln N - N\sett(\eps) + o(N)$, we find that $\sett(\eps)$ is given by
\begin{equation}
\sett(\eps) = \left\{\begin{array}{ll}
 \ln\eps &\mathrm{for}\ \eps \geq 1
\\	\eps-1   &\mathrm{for}\ \eps < 1
\end{array}\right..
\end{equation}
	At $\eps\ (:= 4\veps/\gamma) = 1$ the function $\eps\mapsto \sett(\eps)$ is continuous together 
with its first derivative, but its second derivative is not, giving the signature of a 
second-order phase transition in the sense of Ehrenfest, associated with a symmetry-breaking bifurcation.
	We remark that by evaluating the dispersion-relation obtained from the Jeans (a.k.a. Vlasov-Poisson) 
equations for perturbations about the uniform perfect gas state one directly 
finds that the second-order phase transition point is determined by the Jeans criterion \cite{Jeans,KieAAM} 
adapted to our two-dimensional spherical toy universe.

 Fig.~1 (adapted from \cite{KieJSTAT}) displays $s=\sett(\eps)$ versus $\eps$:
\begin{figure}[H]
\centering
\includegraphics[scale=0.5]{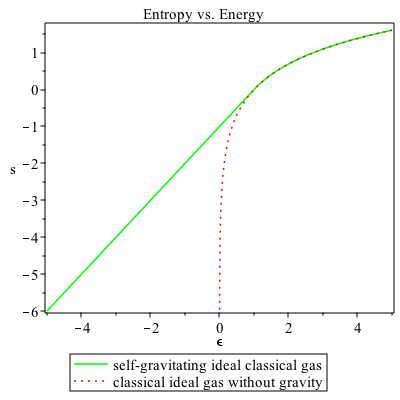}
\label{SvsEin2d}
\end{figure}

 \emph{Conclusions}: 
Clausius' two laws of the universe 1) and 2) are mathematically meaningful for this two-dimensional classical toy universe.
They predict its fate to be a ``heat death'' where all macroscopic evolution ceases.
 In this Hamiltonian system it might do so forever in the limit $N\to\infty$, though
only for a finite (but unpronouncably large) time span of the order of the Poincar\'e recurrence time when $N<\infty$. 
 When $\eps\geq 1$, the ``state of heat death'' looks precisely as envisioned in the earliest papers on this subject: 
a spatially uniform distribution of matter with Maxwellian momentum distribution. 
 However, when $\eps<1$ the ``state of heat death'' is spatially non-uniform, showing gravitational condensation in one 
hemisphere and rarification in the other, yet rotational symmetry about some arbitrary axis $\aV$ (which presumably is 
determined by the initial data).
 Since for $0<\eps<1$ the spatially uniform Maxwellian is also a solution to the Euler--Lagrange equations of the maximum 
entropy principle, to anyone who has learned that an increase in entropy means a decrease in structure it may seem paradoxical 
that the spatially uniform Maxwellian doesn't have the largest entropy.
 Yet note that the gravitationally condensed states have a higher temperature than the uniform Maxwellian state of the 
ideal gas with same energy $\eps$ (or $\veps$).
 In fact the decrease of entropy due to the increase in spatial structure inflicted by gravity is 
overcompensated by an increase of entropy due to the accompanying decrease of structure in momentum space --- 
paradox resolved.

 \textit{First comment}: The reason for why an unbounded entropy does not feature in our model world on $\Sset^2_R$ 
has nothing to do with the dimensionality of the space itself, but with the strength of the singularity of
the gravitational pair interactions in these different-dimensional worlds. 
 Thus, replacing $\frac{1}{\chord{\qV-\qV'}}$ by $\ln\frac{1}{\chord{\qV-\qV'}}$ in the $(\Sset^3_R)^N$ integral in 
(\ref{SBergoINT}) renders (\ref{SBergoINT}) finite for all $N$ \cite{KiePHYSICA}.

 \emph{Final comment}: In \cite{KieJSTAT} also the asymptotic large $N$ expansion of the entropy of the 
maximum entropy state with prescribed energy (scaling like $E=N^2\veps\in\Rset$) and prescribed angular momentum
(scaling like $\LV = N^{3/2}\lambdaV\in\Rset^3$) is studied. 
 In this case the maximum entropy state is spatially non-uniform whenever $\lambdaV\neq 0$; moreover, when $\lambdaV\neq 0$
this state is also not static but either stationary (exhibiting stationary flow) or (quasi-) periodic in time --- see
\cite{KieASSISI}.
 For large $\veps$  the entropy maximizer is rotation-symmetric about $\lambdaV$, but for sufficiently 
low $\veps$ it's not. 
 However, in contrast to the ergodic ensemble (i.e., ignoring angular momentum as constant of motion), 
the maximum entropy principle of this ergodic subensemble does not seem to be 
explicitly solvable in terms of known functions, and the complete classification 
of all maximum entropy states is open.

\newpage

%

\end{document}